\documentclass{aa}
\usepackage{graphicx}
\newcommand{\corot}{{\textsc{CoRoT}}}
\newcommand{\ind}[1]{_{\mathrm{#1}}}
\newcommand{\lat}{\lambda}
\newcommand{\incl}{i}
\newcommand{\Teq}{T\ind{eq}}
\newcommand{\Tmoy}{\overline{T}}
\newcommand{\tm}{t\ind{max}}
\newcommand{\ntache}{N\ind{s}}
\newcommand{\nmoy}{n}
\newcommand{\vie}{\tau}
\newcommand{\duree}{{\mathcal{D}}}
\newcommand{\contrast}{{\mathcal{C}}}
\newcommand{\diff}{\mathrm{d}}
\newcommand{\etaHF}{\eta\ind{32\,s}}
\newcommand{\etaBF}{\eta\ind{orbit}}
\newcommand{\setp}{\mathcal{S}}

\def\muHz{\,$\mu$Hz}         
\def\vsini{v\sin i}          
\def\Phiref{\Phi_0}
\begin{document}
\title{Short-lived spots in solar-like stars as observed by CoRoT~\thanks{The CoRoT space mission, launched on 2006 December 27, was
developed and is operated by the CNES, with participation of the
Science Programs of ESA, ESA's RSSD, Austria, Belgium, Brazil,
Germany and Spain.}}
\titlerunning{Short-lived spots in solar-like stars}
\author{%
B. Mosser\inst{1}\and
F. Baudin\inst{2}\and
A.F. Lanza\inst{3}\and
J.C. Hulot\inst{2}\and
C. Catala\inst{1}\and
A. Baglin\inst{1}\and
M. Auvergne\inst{1}
}
\offprints{B. Mosser}
\institute{LESIA, CNRS, Universit\'e Pierre et Marie Curie, Universit\'e Denis Diderot, Observatoire de Paris, 92195 Meudon cedex, France\\
\email{benoit.mosser@obspm.fr}
\and
Institut d'Astrophysique Spatiale, UMR8617, Universit\'e Paris XI, B\^atiment 121, 91405 Orsay Cedex, France
\and
INAF-Osservatorio Astrofisico di Catania, Via S. Sofia, 78, 95123 Catania, Italy
}
\date{Accepted in A\&A, special volume on first CoRoT data}

\abstract{CoRoT light curves have an unprecedented photometric quality, having simultaneously a high signal-to-noise ratio, a long time span and a nearly continuous duty-cycle.}
{We analyse the light-curves of four bright targets observed in the seismology field and study short-lived small spots in solar-like stars.}%
{We present a simple spot modeling by iterative analysis. Its ability to extract relevant parameters is ensured by implementing relaxation steps to avoid convergence to local minima of the sum of the residuals between observations and modeling. The use of Monte-Carlo simulations allows us to estimate the performance of the fits.}%
{Our starspot modeling gives a representation of the spots on these stars in agreement with other well tested methods.
Within this framework, parameters such as rigid-body rotation and spot lifetimes seem to be precisely determined. Then, the lifetime/rotation period  ratios are in the range 0.5 - 2, and there is clear evidence for differential rotation. }%
{}

\maketitle

\section{Introduction\label{introduction}}

Analyzing stellar activity is important for many topics. Stellar activity has been studied to increase the efficiency of the detection of Earth-like planetary transits  (\cite{2008A&A...482..341B}). Tracers of activity can be monitored with different techniques. White light photometry is not the most sophisticated method, and is affected by many well known drawbacks (e.g., poor determination of the stellar inclination, non-uniqueness of the solution). However, this method is being revised by new space data, as provided by the MOST satellite (\cite{2003PASP..115.1023W}) and now by CoRoT. The space mission CoRoT provides us with 2 years of low noise, long duration, high duty cycle, continuous time series. The quality of the data, compared to previous results, allows us to investigate the activity of solar-like stars, and forces us to revise the performance of white light photometry.

The review paper by \cite{2002AN....323..336C} summarizes the method for observing differential rotation, with a clear focus on the development of the Doppler imaging technique. Results, due to the high capability of this instrumental technique, were obtained on rapidly rotating stars (\cite{2007AN....328.1030C}), such as the determination of the magnetic map of a rapidly rotating, very-low-mass, fully convective dwarfs through tomographic imaging from time series of spectropolarimetric data (\cite{2006Sci...311..633D}), or observations of slowly-rotating but active giants (\cite{2008A&A...491..499A}). Applying tomographic imaging techniques requires a study of active stars with a magnetic field strong enough to produce detectable circularly polarized signatures (\cite{2008MNRAS.390..567M}).

Space-borne precise photometry opens new windows, as already shown by MOST observations.
Differential rotation has been detected by MOST in the light curve of $\varepsilon$ Eridani, a young Sun-like star hosting a planetary companion (\cite{2006ApJ...648..607C}),
and in $\kappa^{1}$ Ceti (\cite{2007ApJ...659.1611W}). Both targets have magnitude variations of about 0.01. With \corot, it becomes possible to investigate in great detail activity in solar-like stars with much smaller variations.
It is for example possible to examine a star such as HD 49933, that clearly shows a signal modulated by rotation (\cite{2005A&A...431L..13M}, \cite{2008A&A...488..705A}), but no spectropolarimetric signature (Catala, private communication).

This work addresses the study of the light-curves of bright stars with short-lived spots. In most cases, these stars exhibit small spots, comparable to the solar spots at maximum activity. We have limited the analysis to the bright solar-like targets observed in the seismology field. For these targets, we have investigated the possibility of deriving precise information on the stellar rotation, including possible differential rotation, spot lifetimes and stellar inclination. As was done for the first analysis of the MOST data, we use the analytic model of \cite{1987ApJ...320..756D} to model the signature of the spots and investigate the case of small variations due to many short-lived spots.

Observations and light curves are presented in Sect.~\ref{observations}. The relevant star and spot parameters for the modeling are explored in Sect.~\ref{method}. The performance we may achieve, derived from the analysis of synthetic time series and hare-and-hounds exercises, is given in Sect.~\ref{performance}. Section~\ref{discussion} is devoted to results on the CoRoT time series and to discussions, and Sect.~\ref{conclusion} to conclusions.

\begin{table*}
\caption{Targets and time series properties.}\label{time-serie}
\begin{tabular}{rccccccccc}
  \hline
  star & type & $m\ind{V}$ &\multicolumn{2}{c}{beginning}& length& $\etaHF$ & $N\ind{inter}$ & $\etaBF$  \\
       &      &            & (date)& (\corot\ day)& (day) & (\%) &  & (\%) \\
\hline
HD 49933  & F5V& 5.78 &31 jan 07& 2587.0&  60.7 & 93.1 & 0 & 100  \\
HD 175726 & G0V& 6.72 &11 apr 07& 2657.2& 27.2  & 90.2 & 0 & 100 \\
HD 181420 & F2V& 6.57 &11 may 07& 2687.1& 156.6 & 91.7 & 3 & 99.86 \\
HD 181906 & F8V& 7.6  &11 may 07& 2687.1& 156.6 & 91.4 & 3 & 99.86 \\
  \hline
\end{tabular}
\end{table*}

\section{Light curves\label{observations}}

We chose to focus this analysis on the main targets of the CoRoT seismology field already observed:
HD 49933 (\cite{2008A&A...488..705A}),
HD 175726 (\cite{mosser2009}),
HD 181420 (\cite{barban2009}) and
HD 181906 (\cite{garcia2009}).
The list of the stars and their characteristics are presented in Table \ref{time-serie}.
We worked with the `Level 2' light curves, ready for the scientific analysis, delivered by the CoRoT pipeline after nominal corrections  (\cite{2007astro.ph..3354S}). For spot modeling, the 32-s initial sampling of the time series is much too dense, so that we rebin the data to one point per CoRoT orbit (about 6184\,s). With this sampling time, the time coverage provided by \corot\ gives a nearly 100\,\% duty cycle, except for very short periods when data were lost due to telemetric problems between the satellite and the ground antennas.
This resampling is coherent with our spot modeling : it remains tight enough compared to the stellar rotation period and has no influence on the spot map since the relevance of the fitting process exposed in paragraph \ref{life-number} requires the number of spots to be limited.

Table \ref{time-serie} gives the value of the duty cycles $\etaHF$ and $\etaBF$, respectively calculated with a rapid or a slow sampling. The duty cycle $\etaBF$ useful for the spot analysis depends mainly on the number $N\ind{inter}$ of long interruptions (longer than 1 CoRoT orbital period) in the data transfer from the satellite to the ground. This number is in all cases less than or equal to 3.

The signal-to-noise ratio gains much with such an undersampling compared to the initial data. The relative standard deviation of the light curves due to the photon noise is then about 10 ppm for the considered targets, much below the intrinsic stellar variations.

The relative flux variations were then calculated assuming a reference level $\Phiref$ close to the observed maximum stellar flux. In order to remove the non-stellar drifts related to the ageing of the CDD, we had to remove a polynomial term, close to the linear trend characterized by  $\diff \ln \Phi/ \diff t =-5.32\ 10^{-5}$ d$^{-1}$ reported by
\cite{2009arXiv0901.2206A}. The maximum degree of the correction takes into account the part of the variation that is correlated for all light curves of a given run.

\begin{figure}
\centering
\includegraphics[width=8.5cm]{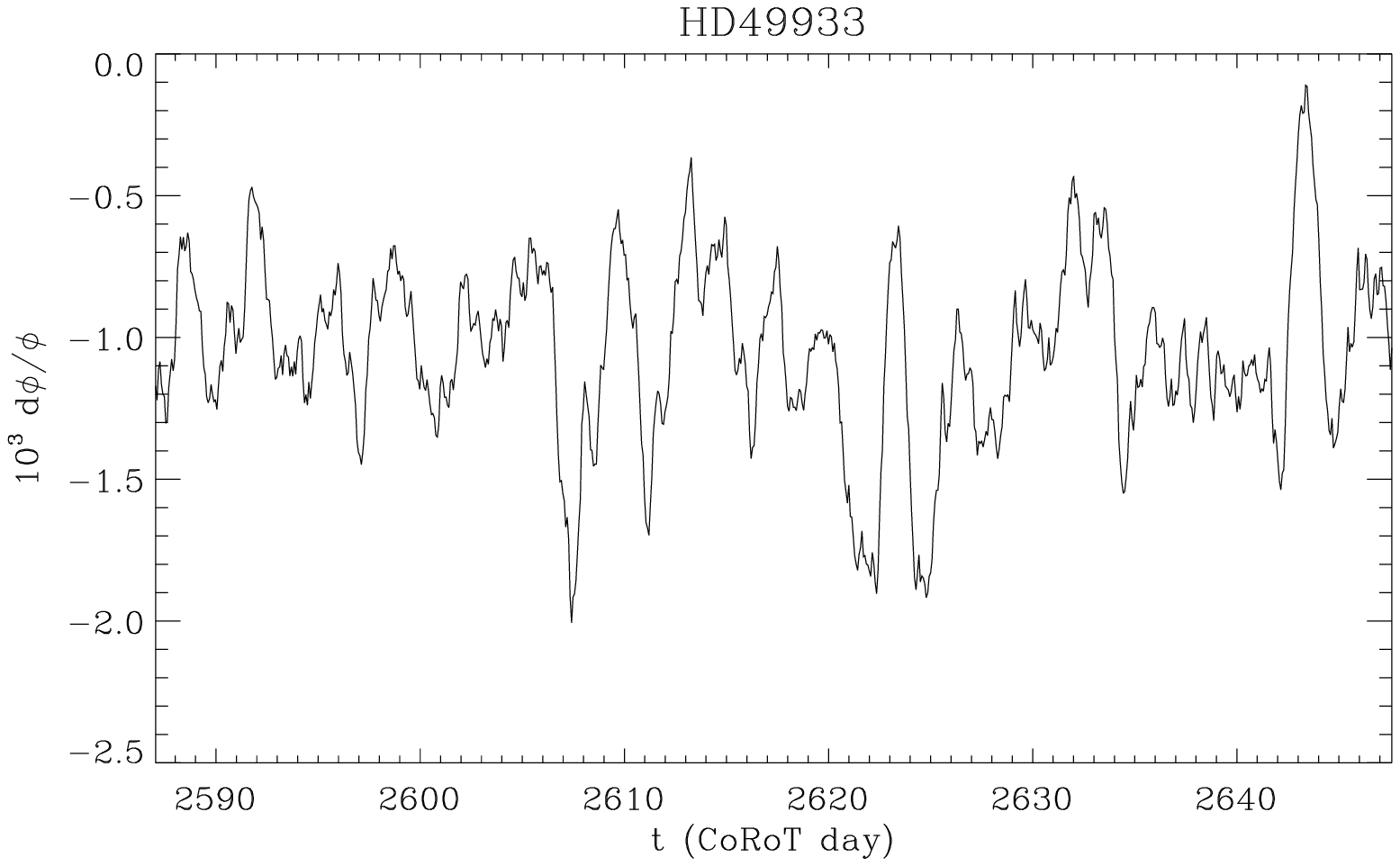}
\includegraphics[width=8.5cm]{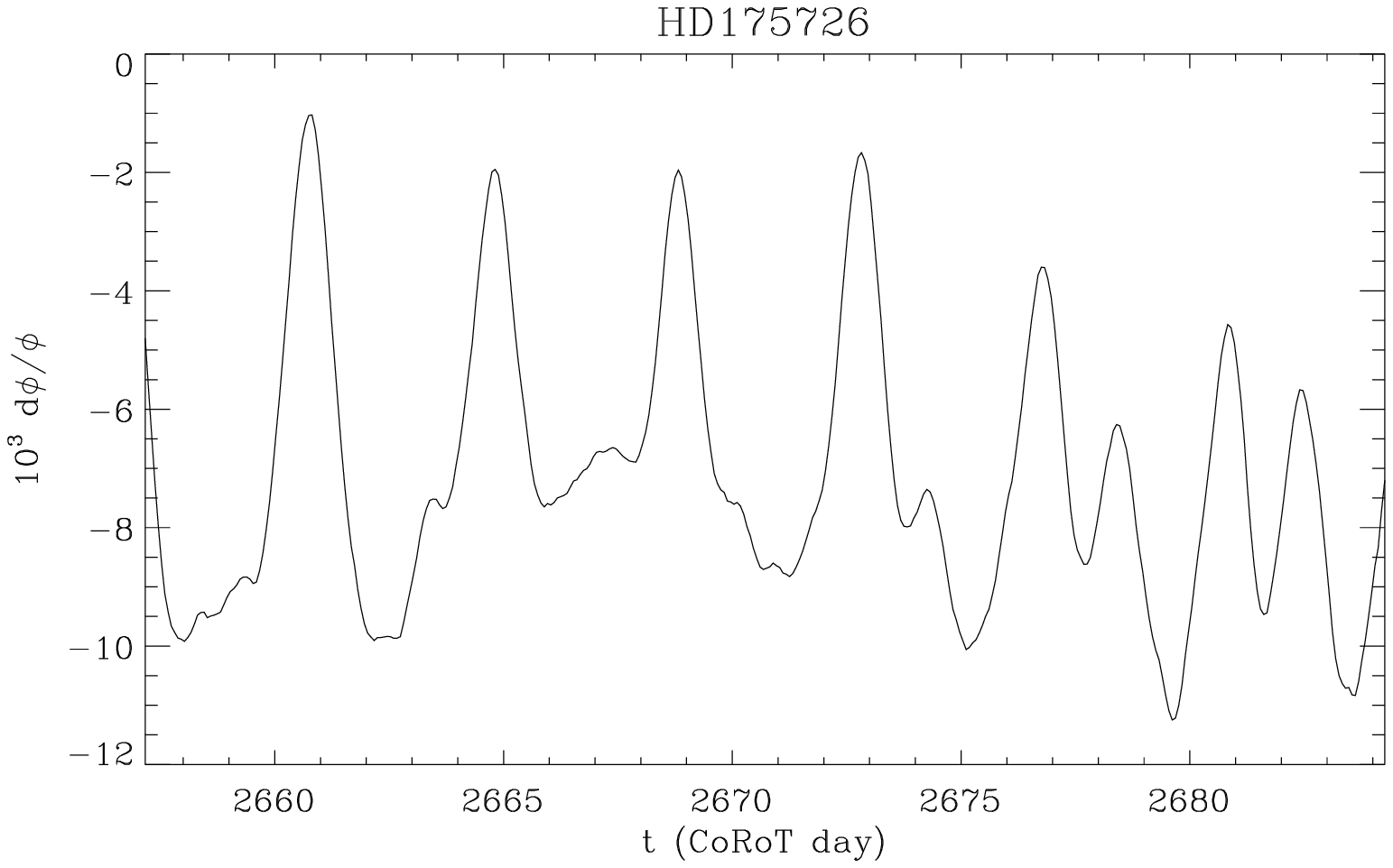}
\includegraphics[width=8.5cm]{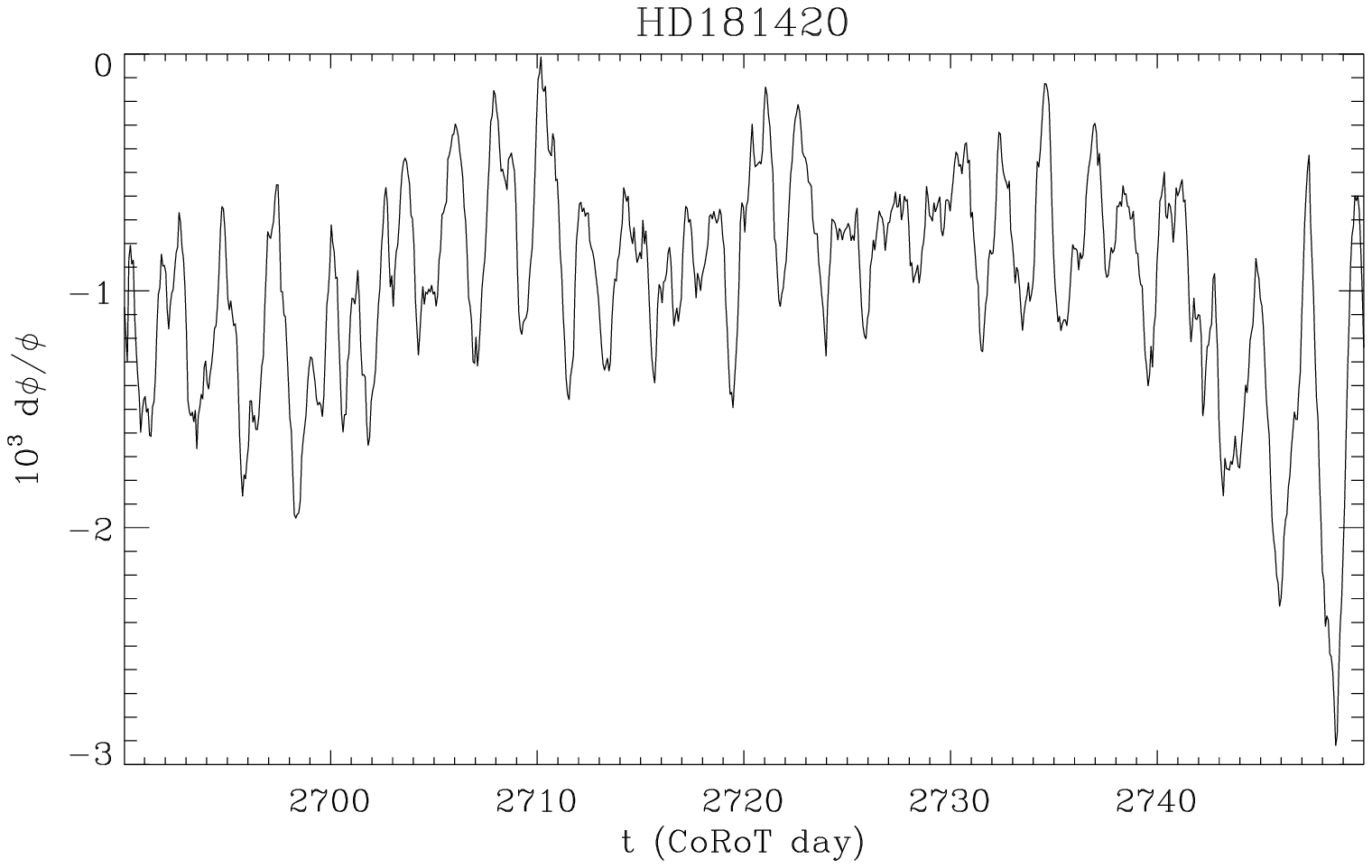}
\includegraphics[width=8.5cm]{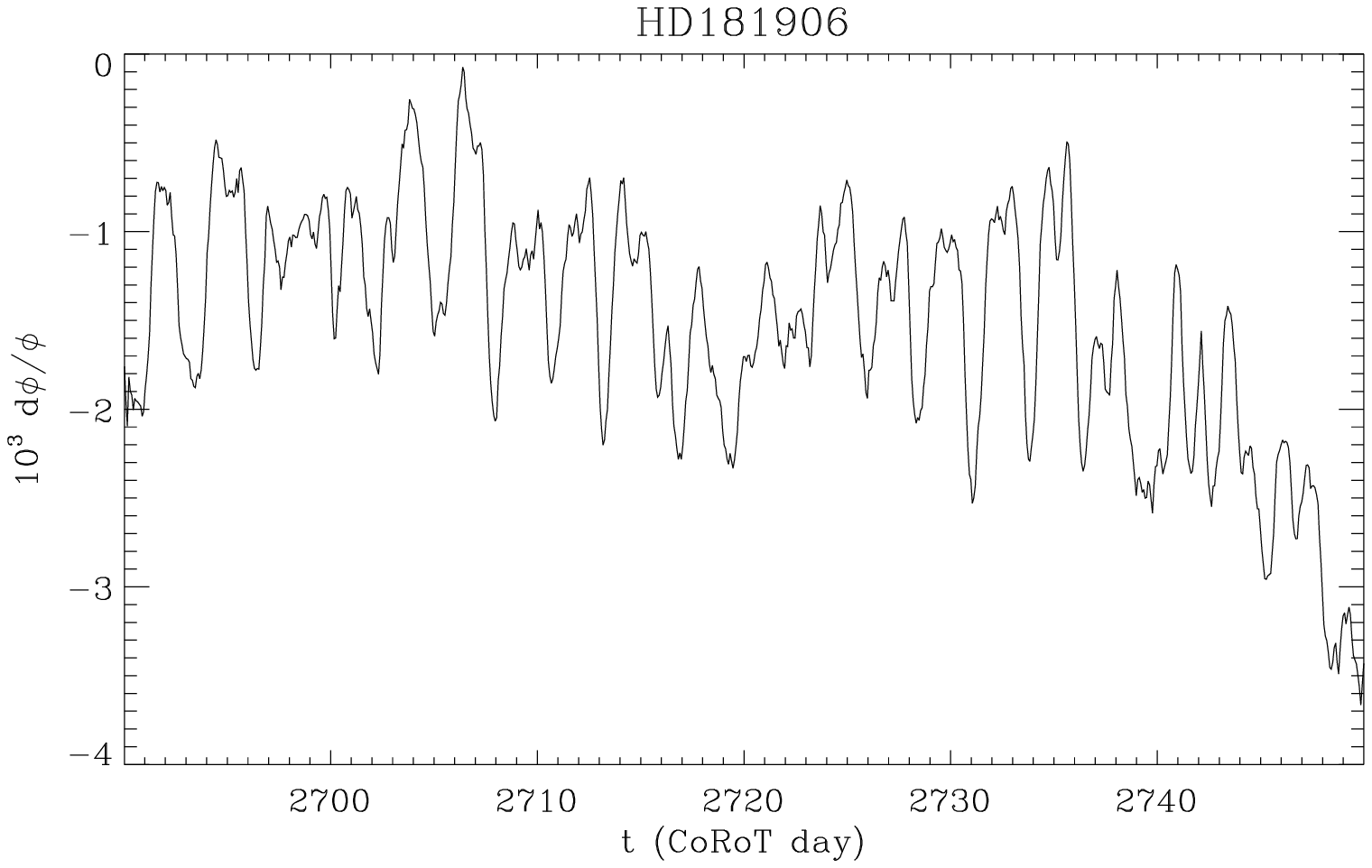}
\caption{Time series of the relative flux variations of 4 CoRoT bright targets observed in the seismology field. Time sampling has been reduced to 1 point per CoRoT orbit (6184\,s). Only a 60-day sample is shown for HD 181420 and HD 181906.
\label{timeseries}}
\end{figure}

We chose to analyze 60-day long time series (except for HD175726, observed for 27.2 days during a short run).
For stars observed during the longer than 120-day runs, we had then the possibility to split the series and compare completely independent subseries.
Figure~\ref{timeseries} shows the light curves. Except for HD 175726, the typical relative variations, of about 2 to 3\,mmag, are similar to the active Sun (\cite{2004A&A...414.1139A}).

\section{Modeling\label{method}}

The form of the time series indicates that the light curve is modulated by a large number of small short-lived spots. We have chosen to explicitly address this case, with a simple model  following the formalism developed by \cite{1987ApJ...320..756D}, modeling the colder stellar patches by individual circular spots. We do not introduce either penumbra  or facula. As in the data observed with MOST and recently analyzed, we have to specifically solve for the elements of inclination, differential rotation and  equatorial rotation period.

The parameters of interest are, for the star, its inclination angle $i$, the unspotted stellar flux $\Phiref$, the limb darkening parameters and the temperature contrast between the spots and the stellar photosphere.  A spot is characterized by the set of an epoch for the maximum spot intensity, a longitude, a latitude, and an angular size.
Rotation and lifetime of the spots have an intermediate status: depending on the analysis, they were set to the same value for the whole star or left free for each spot. The stellar global parameters considered as fixed inputs during each spot modeling were tested on extensive grids. A large set of values were tested, so that their possible variations were precisely analyzed.

\subsection{Limb darkening; spot temperature }

The linear limb darkening function introduced in the model is derived from \cite{1995A&AS..110..329D}, according to each stellar type.  We do not study the influence of the limb darkening parameters on the fitting process, since the problem in white light is highly degenerate. Due to the fact that the spots for the considered set of stars are small, with a radius of a few degrees or less, they behave essentially as points. Then, a larger spot radius will compensate for an underestimated spot temperature deficit, and vice versa. Moreover, white light photometry is not suitable for a precise determination of those parameters.

\subsection{Unspotted flux, unspotted area}

The value of the unspotted stellar flux is an important issue for this study. A too high unspotted level yields for example overestimated spot latitudes, and inversely. Its determination therefore has a large incidence on all parameters sensitive to the spot latitude; we verified that the incidence on the mean rotation and the spot lifetime is limited.

There is unfortunately no way to obtain an estimate of this parameter, and it is not possible to disentangle it from the flux variations due to the aging of the CCD camera or due to the perturbations caused by high-energy particles impacting the CCD. Improving this point will require better knowledge of the CoRoT ultimate photometric precision for long-term variation. This work will be undertaken in a further work dedicated to differential rotation.

\subsection{Lifetime and number of spots\label{life-number}}

First simulations have verified that, as in the solar case, the lifetimes of the small stellar spots are short, and comparable to the stellar rotation period. We chose to model the time evolution of the spot contrast $\contrast$ according to a Gaussian law:
\begin{equation}
\contrast (t) = \contrast\ind{max}\  \exp{ \left[ - \left(\ln 2\, {t-\tm \over \vie} \right)^2\; \right] }
\label{vie}
\end{equation}
with $\tm$ the epoch at maximum contrast $\contrast\ind{max}$, and $\tau$ the lifetime.
All spot contributions construct the flux variation:
\begin{equation}
\Phi (t) = \Phiref\  \left[ 1 - \sum_{i=1}^{\ntache} \contrast_i (t)\right]
\label{dflux}
\end{equation}
where $\Phiref$ is the unspotted flux introduced in Section~\ref{observations}.
The total number of spots $\ntache$ during a lapse of time $\duree$ has been chosen so that there are on average $\nmoy$ spots per rotation. This is ensured by the relation:
\begin{equation}
\ntache\ \vie =  \nmoy\ \duree
\end{equation}
The spot density $\nmoy$ is about 2 to 3 per rotation period. A larger number of spots reduces the residuals of the best fit, but
degrades its significance, by introducing, for example, large groups of close spots, whose parameters are highly degenerate, hence imprecise.

\subsection{Stellar inclination and spot latitudes}

The information on latitude derives from the fact that the transit time of one spot depends mainly on this parameter. Assuming that, due to possible degeneracies, the time of visibility is the only source of uncertainty, an uncertainty $\delta \incl$ on the inclination $\incl$ translates directly into an uncertainty $\delta\lat$ on the latitude  $\lat$ according to the conservation of the ratio:
\begin{equation}
{\tan \lat \over \tan \incl} =  {\tan (\lat + \delta\lat) \over \tan (\incl+ \delta \incl)}
\label{uncertainty}
\end{equation}
As a result, underestimating the inclination leads to an underestimation of the spot latitudes, and vice versa. Even if the way the fitting process determines the latitude is more complex due to possible interactions between close spots, relation (\ref{uncertainty}) is useful in estimating the uncertainty: a 10$^\circ$ uncertainty on the star inclination may translate into a much higher uncertainty on the spot latitudes, which may hamper any precise determination of the differential rotation rate of the star.

\begin{table}
\caption{Different fitting methods tested, with fixed or free parameters. Free parameters are fitted; fixed parameters are tested on a grid of values. \label{methode}}
\begin{tabular}{lllll}
  \hline
Fitting method & A & B & C & D\\

\hline
inclination $i$          & fixed & fixed & fixed & fixed \\
number of spots $\ntache$& fixed & fixed & fixed & fixed \\
lifetime                 & free  & fixed & fixed & fixed \\
rotation                 & free  & free  & fixed & fixed \\
differential rotation    & --    & --    & 0     & fixed \\
spots                    & free  & free  & free  & free  \\
  \hline
stellar parameters       &     2 &    3  &     4 &   5   \\
spot parameters          &     6 &    5  &     4 &   4   \\
parameters  &2+6$\ntache$&3+5$\ntache$&4+4$\ntache$&4+4$\ntache$\\
\hline
\end{tabular}\end{table}

\section{Fit and performance: hare-and-hounds exercises\label{performance}}

Fitting was performed by an iterative linear search first, followed by a Powell search
(\cite{powell}), the $\chi^2$ function to be minimized being the sum of the squared residuals. Measurements uncertainties being nearly constant due to the high photometric stability provided by \corot, the reduced $\chi^2$ are simply obtained by dividing the $\chi^2$ value by a constant term corresponding to the measurement uncertainty. As usual in photometric spot modeling, these reduced $\chi^2$ are high. During an iteration step, all spot parameters have their values fixed by the preceding step, but the spot whose parameters are fitted.  The extent of the exploration of these parameters decreases during the iterative process.

Solutions are relaxed to prevent the solution from stalling in a local minimum and to quantify the non-uniqueness problem. During the first steps of the iteration, relaxation consists of resetting the spot latitudes to zero or in adding large perturbations on all spots parameters. In the following steps, we control the extent of the exploration of the parameters {in order to simultaneously enhance the precision and reach the convergence as rapidly as possible}. Globally, changes become increasingly smaller, but similarly to simulated annealing, the control includes random changes. These relaxation steps are achieved automatically during the fitting process, according to a global empirical scheme optimized when treating the hare-and-hounds exercises. This scheme defines the number of iterations with a given step, the criterion for allowing the next step, the decrease of the changes of the exploration, and the occurrence of the relaxation events.
During the fitting process, spots can be removed when their contribution becomes negligible, due to a too small angular radius and/or a too high latitude.

With 4, 5 or 6 free parameters for each spot, depending on whether the rotation and lifetime are fixed or not (Table~\ref{methode}), and typically 50 spots for each 60-day long sequence, the number of the parameters to be fitted varies from about 200 to 300. Fitting method A, with both spot lifetimes and spot rotations considered as free parameters, was used as a first step. This method A is useful for determining the relevant ranges of these parameters, hence for defining the grid of the values tested by the other fitting methods. It shows that spot lifetimes are mainly distributed in a restricted range. Therefore, the lifetime $\tau$ was then considered as a fixed stellar parameter for all other fitting processes. Spot rotation period is free in fitting method B, and fixed to a single value $\Tmoy$ in method C. Differential rotation is fixed by 2 parameters in method D, according to the rotational law:
\begin{equation}\label{difrot}
T(\lat) = {\Teq \over 1 - K\sin^2 \lat}
\end{equation}
with $\Teq$ the equatorial rotation period and $K$ the differential rotation rate.

Before applying the spot modeling to the CoRoT targets, we have tested it on synthetic light curves. An extensive study has been provided, with blind analysis. Synthetic light curves were constructed by one of us and given to the analyzer without any information. The parameters of the synthetic light curves were varied in order to mimic the variety of the parameters estimated in the real stars. In order to cover a set of cases at least as large as the observations, we tested inclinations in the range [10 - 90$^\circ$], lifetimes $\vie$ with ratios $\vie/\Tmoy$ in the range [0.5 - 2], and mean spot radius below 0.5$^\circ$.

The method proved to be robust for retrieving the stellar parameters according to the results of these hare-and-hounds exercises. With numerous tests, we were able to derive realistic uncertainties, derived from the standard deviation of the difference between the input and output values.
These uncertainties on the rotation, spot lifetime and star inclination are presented in the following paragraphs.

\subsection{Rigid-body or differential rotation}

Simulations have shown that the determination of the mean rotation $\Tmoy$ is robust.
The uncertainty we get on this parameter is about 0.5\,\% in most cases, up to 1\,\% in the unfavourable case of very short-lived spots or of stellar inclination close to the pole-on case. Of course, when differential rotation is present, the determination of $\Tmoy$, defined as the mean value of the spot rotations, is possibly affected by the distribution in latitude of the spots. Simulations show that $\Tmoy$ remains however a relevant parameter.

The determination of the differential rotation parameter $K$ is not straightforward. When the star is seen pole-on or edge-on, the degeneracy on the latitudes precludes any reliable determination of the coefficient $K$. Values of about 0, 15 an 30\,\% were tested. Better determinations are found for mid-inclination models. The determination of $K$ is usually underestimated. Precision in the hare-and-hounds exercises is limited to about 70\,\%. A large part of the error comes from the uncertainty on the star inclination.

\begin{table*}
\caption{Output parameters derived from the spot modeling. }\label{results}
\begin{tabular}{rcccccccc}
  \hline
  star &\multicolumn{2}{c}{inclination}& spot contrast& \multicolumn{2}{c}{spot angular radius}& lifetime& mean rotation \\
       & $i\ind{spot}$ &  $i\ind{seismo}$& $\langle\contrast\ind{max}\rangle$&median & range&  $\tau$ & $\Tmoy$ \\
       &\multicolumn{2}{c}{\dotfill\ (deg) \dotfill}&($\,^0\!/_{00}$)&\multicolumn{2}{c}{\dotfill\ (deg) \dotfill}&\multicolumn{2}{c}{\dotfill\ (day) \dotfill}\\
\hline
HD 49933  & $50\pm 25^\circ$& $55\pm10^\circ$ (a)& 1  & 1.8 & $1\to2.5^\circ$& $2.5\to3.5$& 3.45$^{+0.05}_{-0.05}$ \\
HD 175726 & $55\pm 25^\circ$&  --                & 8  & 4.7 & $3  \to7^\circ$& $5  \to  7$       & 3.95$^{+0.1}_{-0.1}$   \\
HD 181420 & $60\pm 25^\circ$& $45\pm 4^\circ$ (b)& 0.8& 1.6 &$0.8 \to2^\circ$& $1.5\to  3$    & 2.25$^{+0.03}_{-0.01}$ \\
HD 181906 & $45\pm 25^\circ$& $24\pm 3^\circ$ (c)& 1  & 1.8 &$1.0 \to 2.5^\circ$& $1.5\to  3$ & 2.71$^{+0.03}_{-0.01}$ \\
  \hline
\end{tabular}
\par
Inclinations derived from the spot model are compared to the seismic values inferred from the fit of the Fourier spectra of p modes: (a) Appourchaux et al. (2008); (b) Barban et al. (2009); (c) Garcia et al. (2009).
\end{table*}

\begin{figure*}[!h]
\centering
\includegraphics[width=17.cm]{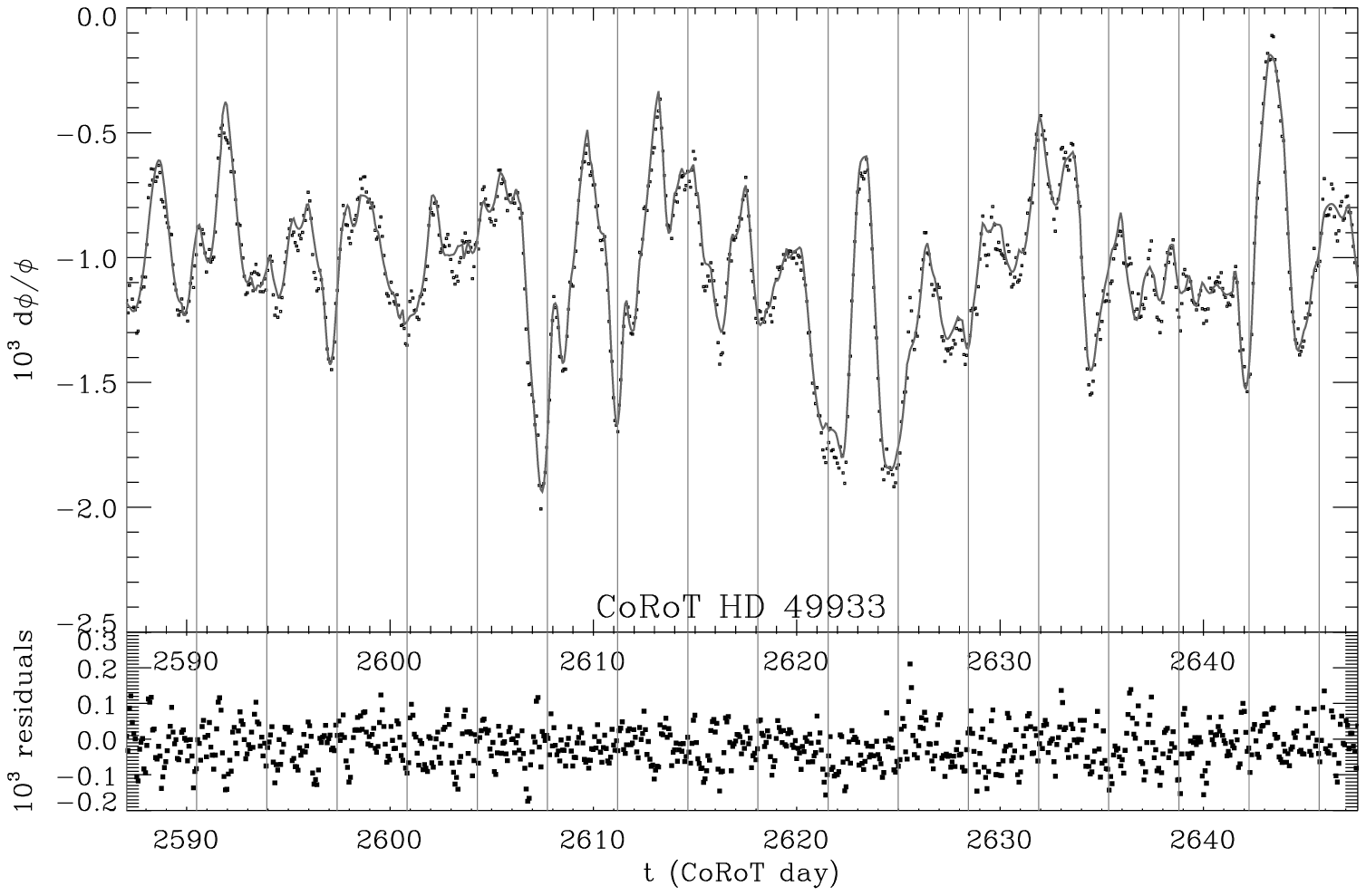}
\includegraphics[width=17.cm]{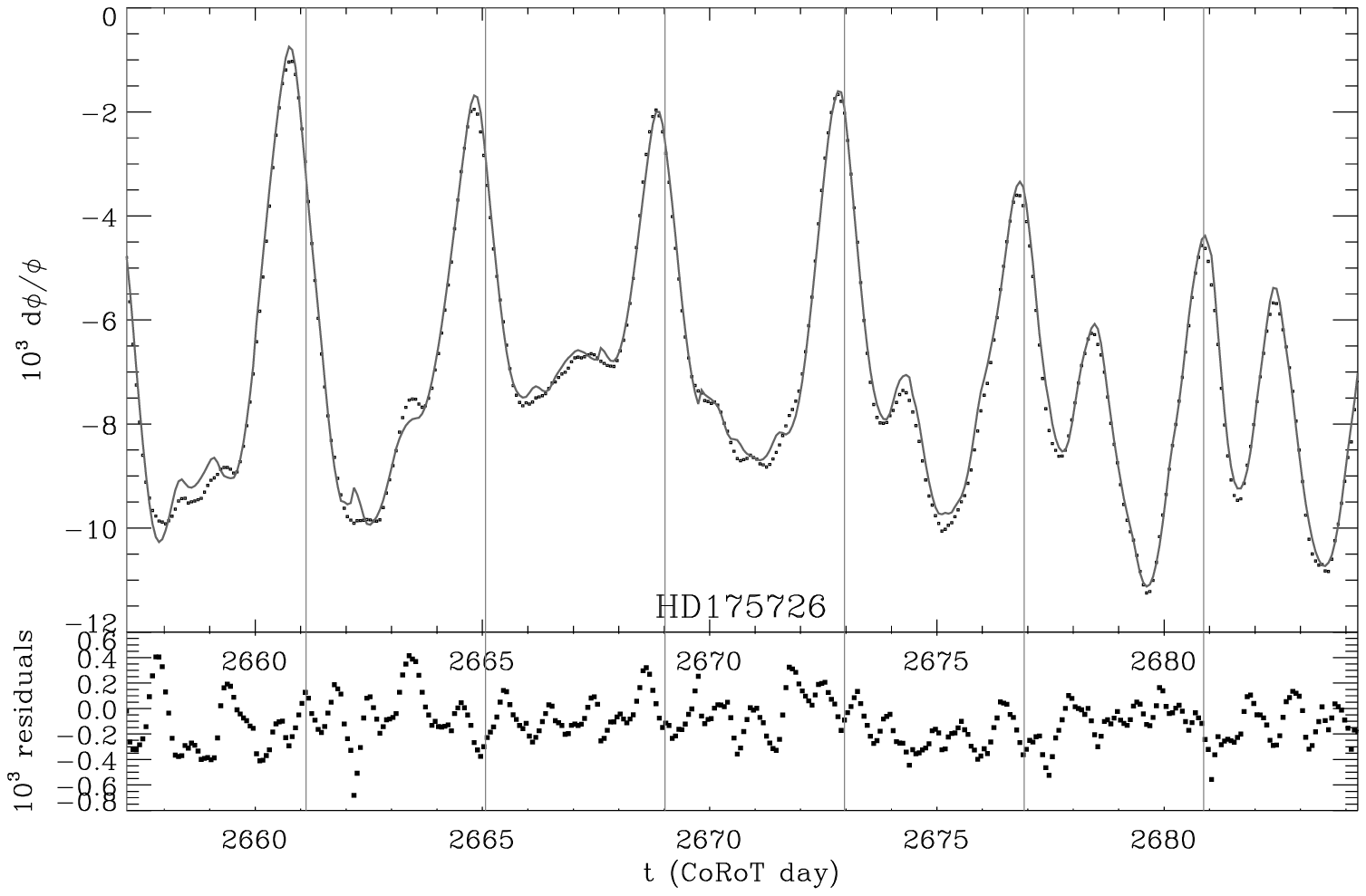}
\caption{Spot modeling of HD49933 and HD 175726. The dots in each upper panel represent the data binned every CoRoT orbit (6184\,s), and the solid curve is the best fit model.  Dots in each bottom panel represent residuals. Vertical grey lines indicate the mean rotation period.
\label{fitcorot}}
\end{figure*}

\begin{figure*}[!h]
\centering
\includegraphics[width=17.cm]{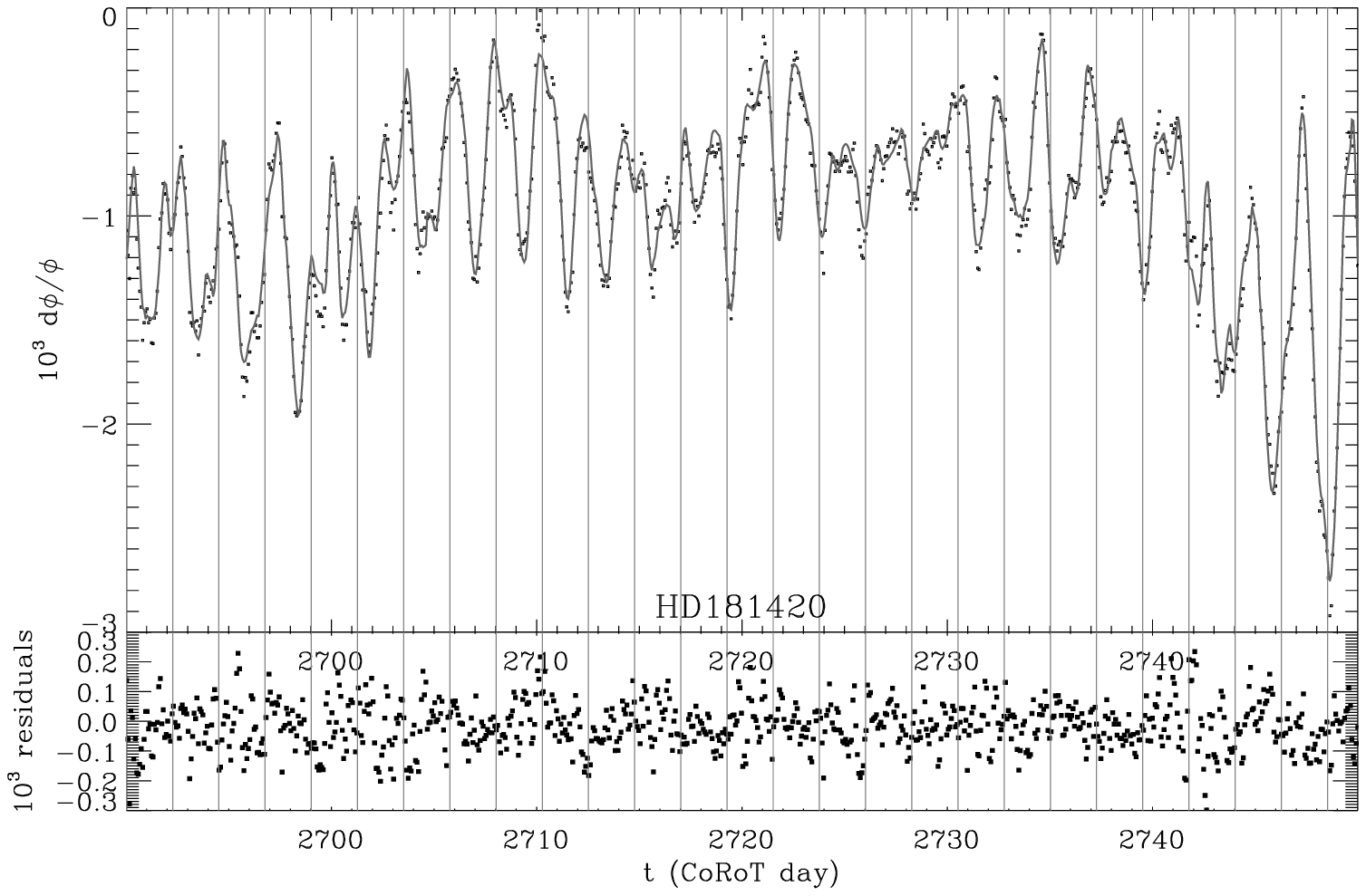}
\includegraphics[width=17.cm]{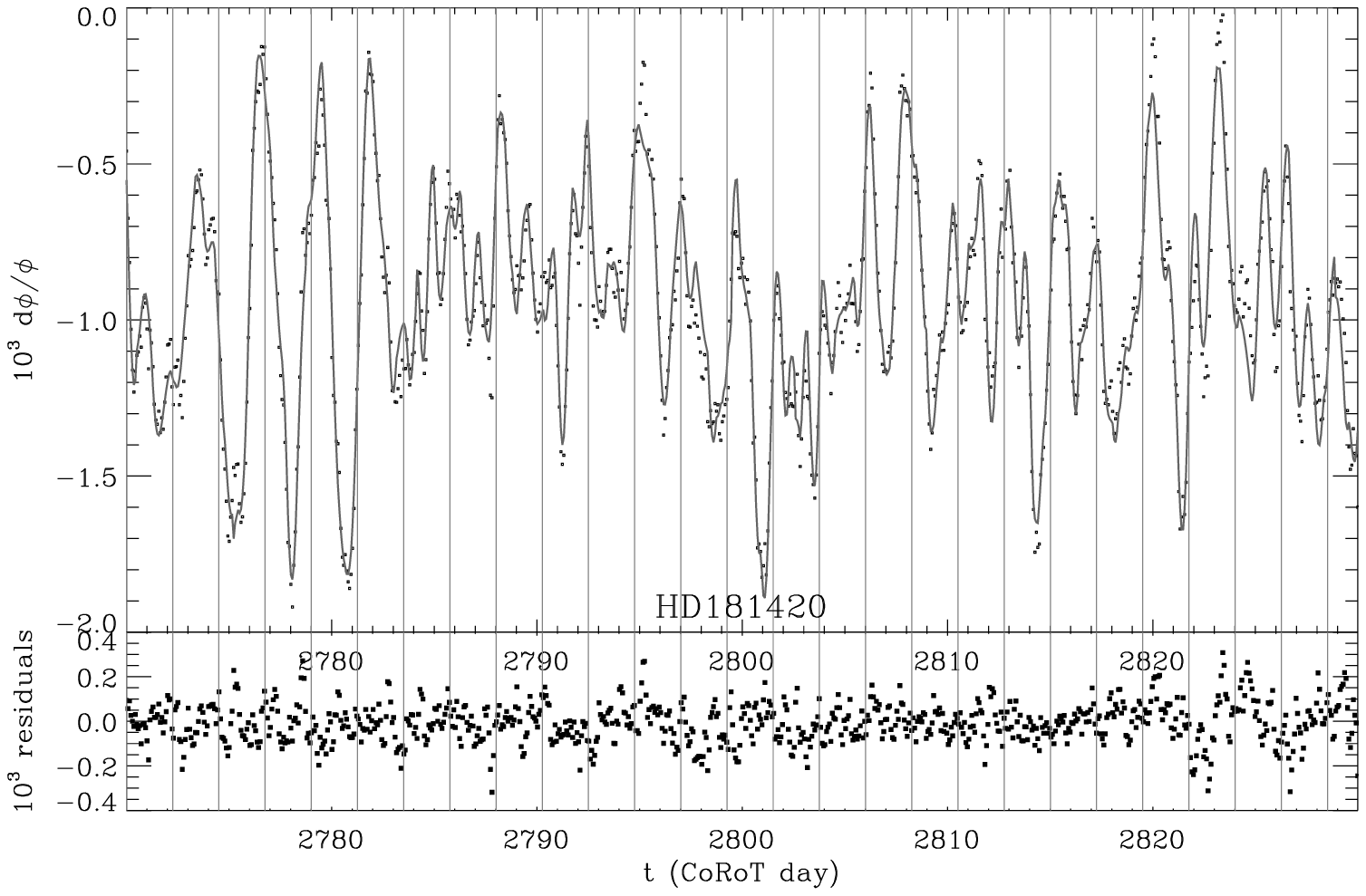}
\caption{Same as Fig.~\ref{fitcorot}, but for HD181420 between CoRoT days 2690 and 2750, and between CoRoT days 2770 and 2830.
\label{fitcorotb}}
\end{figure*}

\begin{figure*}[!h]
\centering
\includegraphics[width=17.cm]{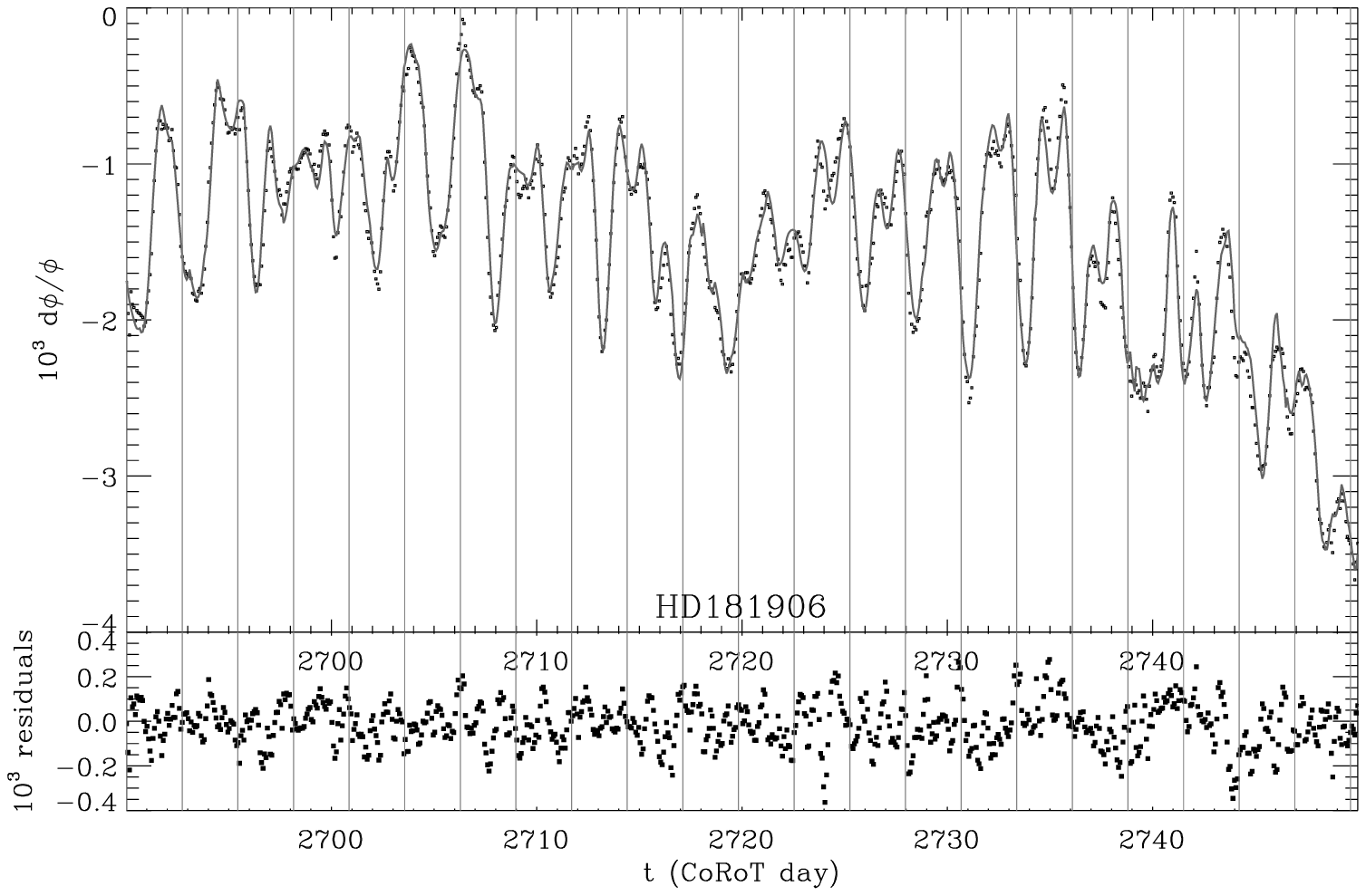}
\includegraphics[width=17.cm]{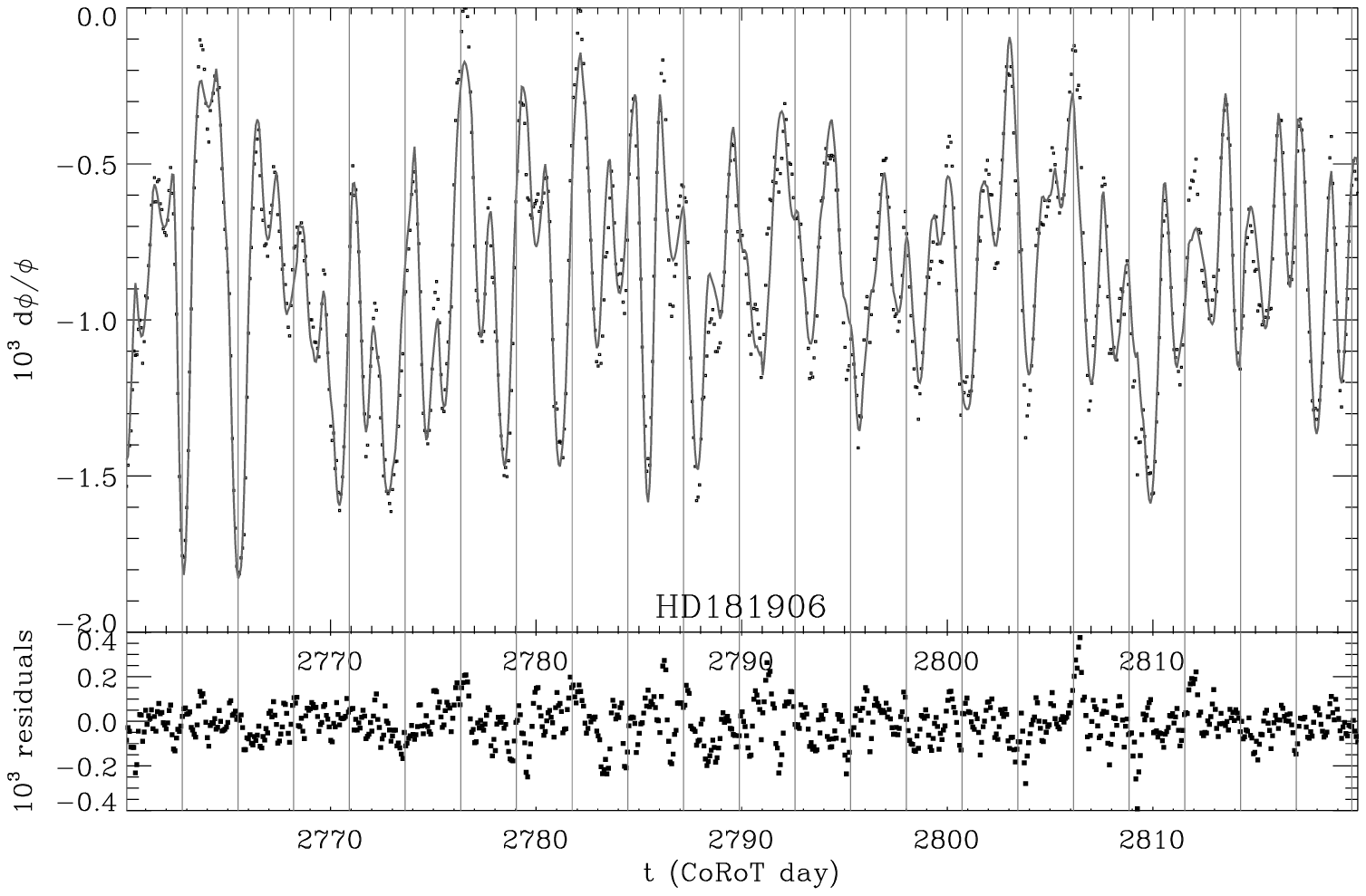}
\caption{Spot modeling of HD181906, same as Fig.~\ref{fitcorotb}.
\label{fitcorotc}}
\end{figure*}

\subsection{Spot lifetime}

As for the rotation, we estimate the performance of the method when trying to retrieve the spot lifetime. The lifetime appears to be slightly overestimated if too few spots were included in the modeling, and underestimated in the opposite case. When the spot lifetime is long compared to rotation, the fitting process is less precise, due to interactions between spots. These interactions may introduce uncontrollable degrees of freedom.

The estimate of the mean spot lifetime $\tau$ in the synthetic time series is robust, and not significantly affected by uncertainties on other parameters. The precision we get on this parameter is about 10\,\%, in the case where the ratio $\vie / \Tmoy$ is in the range [0.5 - 1.5], and about 20\,\% for larger ratios. When lifetimes become much shorter than the rotation period, the error bars on all parameters increase.

\subsection{Star inclination and spot latitudes}

Determining inclinations is well known to be highly difficult with white light photometry. The hare-and-hounds exercises show that it is however possible to recover corrected information on the latitude for intermediate latitudes. Determining the latitude for pole-on on edge-on cases is however very imprecise, due to degeneracy.

To fit the inclination, different values were exhaustively tested between 20 and 90$^\circ$.
The precision on the latitude in the synthetic time series is about $\pm 10^\circ$ for inclinations in the range [30 - 60$^\circ$], increasing to $\pm 20^\circ$ in the range [70-90$^\circ$]. When the star is seen edge-on, no information can be extracted to derive the spot latitude, since all spots transit the star in half a rotation period. When the star is seen pole-on, there is no further transit, and smooth modulations are difficult to fit. It is difficult to distinguish a quiet star from an active one seen pole-on. In all cases, the determined stellar inclination angle depends significantly on the assumptions of the model.

\begin{figure}
\centering
\includegraphics[width=8.8cm]{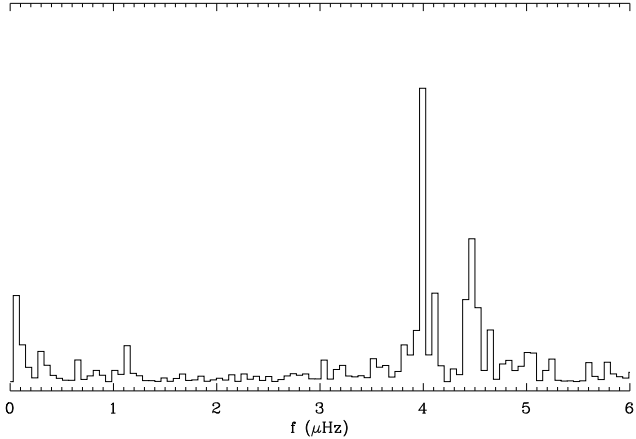}
\caption{Low-frequency spectrum of HD181906. The rotational signature exhibits 2 peaks at about 4.0 and 4.4\,$\mu$Hz (2.9 and 2.6\,d, respectively).
\label{figpics}}
\end{figure}

\begin{figure}
\centering
\includegraphics[width=8.8cm]{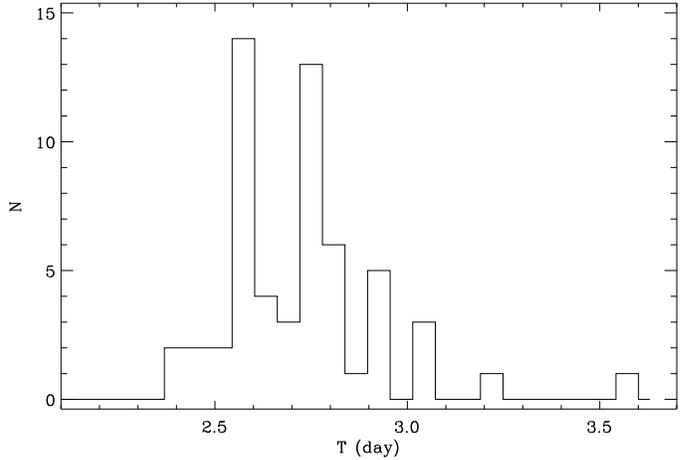}
\caption{Histogram of the distribution of the spot rotations in HD181906.
\label{historot}}
\end{figure}

\section{Results and discussion\label{discussion}}

Hare-and-hound exercises presented above make us confident to analyze the CoRoT time series.
Spot modeling has been applied to 60-day long times series of the CoRoT asteroseismic targets (Fig.~\ref{fitcorot}, \ref{fitcorotb}, \ref{fitcorotc}). The results we obtain for the star inclination, the spot size distribution, the lifetime of the spots and the mean rotation period are presented in Table~\ref{results} and discussed in this Section.
Deriving reliable error bars being not straightforward in the case of a large number of parameters, we applied three different methods.\\
- The tests made on synthetic time series allow us to quantify the goodness of the fit and to extract confidence intervals from the chi-square fitting. Assuming Gaussian statistics, due to the large amount of data, the relative likelihood of a set $\setp$ of parameters can be expressed as \cite{2006PASP..118.1351C}:
\begin{equation}\label{likelihood}
p_\setp \propto \exp\left[ - {(\chi_\setp^2 - \chi\ind{min}^2) \over 2\,T}\right]
\end{equation}
with $\chi\ind{min}^2$ the minimum value reached for the best fit, and $T$ the virtual temperature as used in the case of Markov chain Monte Carlo methods. This virtual temperature should be 1 when the value of the reduced chi-square is about 1. With large values of $\chi\ind{min}^2$, in the range from 50 to 100 in most cases, we should use $T = \chi^2\ind{min}$ as in \cite{2006PASP..118.1351C}. However, hare-and-hounds exercises show that the 1-$\sigma$ uncertainty derived from Eq.~\ref{likelihood} (at 68\,\% of the maximum value reached by $p_\setp$) is then significantly exaggerated. In fact, an irreducible contribution to the $\chi^2$ values comes from high-frequency signatures that the modeling cannot reproduce; it would require the introduction of too many small spots, that would translate into quantitatively good fits, but qualitatively poor, or the introduction of hot faculae. In order to reconcile the performance of the method derived from the hare-and-hounds exercises and Eq.~\ref{likelihood}, we have empirically considered $T\simeq 0.4\,\chi\ind{min}^2$. We are aware that this supposes that the models tested during the exercises were close enough to reality to give reliable indications of the precision.\\
- A second estimate of the confidence interval was achieved by comparing results obtained from tens of fits for each input configuration, derived from different initial conditions and different relaxation tools.\\
- A third estimate was derived from the comparison of independent time subseries, available for HD 181420 and HD 181906. We have verified that the independent best fit values we obtain agree.

Observed error bars were defined in order to encompass all results derived from the fits based on methods B and C, method A being used to define the grid of the parameters to be tested. These error bars are slightly larger than the ones derived from synthetic time series. The partial results of method D are consistent with other methods, but need to be completed by more fits before being reliably presented. This will be done in a forthcoming work, and compared to the results given by the maximum entropy method (Lanza et al. 2009a).

\subsection{Rotation}

The mean star rotation $\Tmoy$ can be precisely inferred from the fits. The relative precision is of about 1\,\% (but 2.5\,\%  for HD 175726, observed for 27.2\,d), comparable to the precision that can be achieved from the study of the low-frequency spectrum (\cite{baudin2009}). The presence of differential rotation is strongly suspected in F dwarfs, as reported by \cite{2003A&A...398..647R}, and, in fact, three out of these four targets show indications of non-solid rotation.

\begin{figure}
\centering
\includegraphics[width=8.8cm]{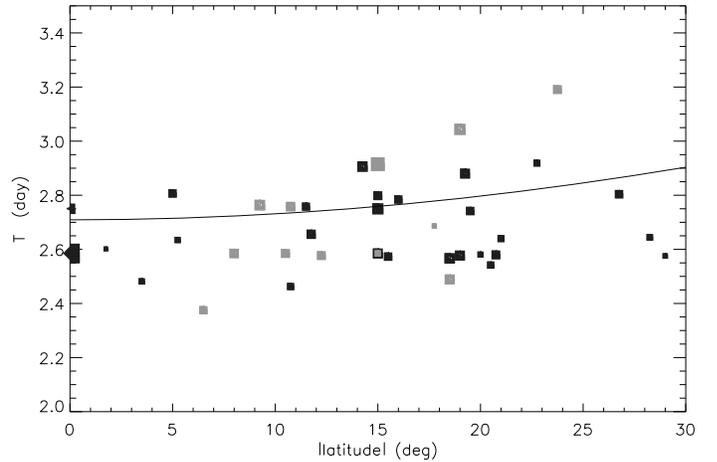}
\caption{Rotation-latitude relation, for HD181906. The size of the solid black or grey squares (for spots  respectively located in the northern or southern hemisphere) is proportional to the spot size. The solid line indicates the best fit of differential rotation according to Eq.~\ref{difrot}.
\label{figpicrot}}
\end{figure}

\begin{figure}
\centering
\includegraphics[width=8.8cm]{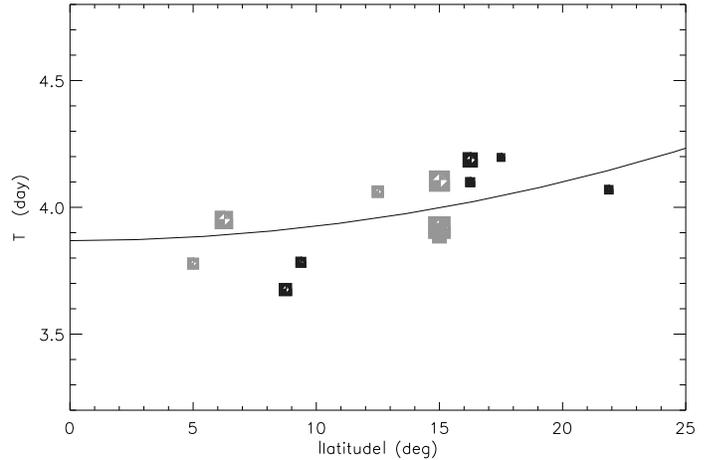}
\caption{Same as Fig.~\ref{figpicrot}, but for HD175726.
\label{figpicrot2}}
\end{figure}

HD 181906 represents the clearest example: as already noticed (\cite{garcia2009}), its low-frequency spectrum (Fig.~\ref{figpics}) exhibit 2 peaks, at about 4.0 and 4.4\muHz. The spot modeling also shows 2 peaks in the histogram of the spot rotations derived from the fitting method B. The 2 peaks at respectively 2.6 and 2.9\,d correspond clearly to the signatures in the Fourier spectrum (Fig.~\ref{historot}). They may be due to a combination of the rotation period and short spot lifetime. We checked that they are more likely due to differential rotation, since the latitudes of the spots are correlated to the rotation rate: the mean latitudes of the spots, sorted according to the rotation period, increase with increasing rotation period (Fig.~\ref{figpicrot}). HD 175726 presents similar features (Fig.~\ref{figpicrot2}).

We can derive the order of magnitude of the differential rotation rate, but with large error bars: $K$ is about 0.25 for HD181906, and about 0.40 for HD175726, in agreement with expected values for such targets (\cite{2006A&A...446..267R}). The case of HD 49933 will be exhaustively considered in a forthcoming work.

\subsection{Spot lifetime}

\begin{figure}
\centering
\includegraphics[width=8.8cm]{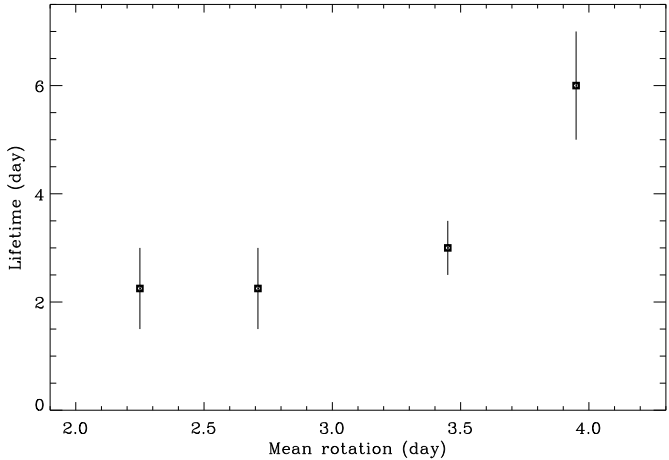}
\caption{Spot lifetime as a function of the mean rotation period.
\label{paramduree}}
\end{figure}

If our spot model is correct, then as expected from the preliminary synthetic analysis,  the determination of the spot lifetimes gives convergent, hence reliable results. The values we obtain fit in the range 0.5$\to$2 times the rotation period. The lifetimes of the spot may increase with increasing rotation period (Fig.~\ref{paramduree}) and with increasing spot size.

The uncertainties we derive on the spot lifetimes vary from $\pm 20$ to $\pm 50$\,\%. Different reasons may explain these large values: the ad-hoc model to describe the evolution with time (Eq.~\ref{vie}), and dependence of the lifetime on the spot size. The study of the low-frequency spectrum of HD 49933 (\cite{baudin2009}) gives similar results, which enhances the confidence we may have in the description of the spot map provided by the model.
We also applied the maximum entropy spot model of \cite{2009A&A...493..193L} to derive information on the active region evolution of HD49933. The time series were subdivided into subsets of one rotation period (i.e., 3.4 days), each of which was fitted with a fixed distribution of spots, optimized to fulfil the maximum entropy regularization criterion. Therefore, this approach is sensitive only to evolutionary timescales longer than one rotation period, in contrast to the discrete spot modeling approach, which can resolve shorter spot lifetime. In spite of such a difference, the results obtained with the two methods are comparable. In the ME model of HD49933, most of the spots have lifetimes between 5 and 7 days, with only $10-15$\,\% showing a lifetime up to $7-10$ days during the short run time series.

\begin{table*}
\caption{Stellar inclinations. }\label{incli}
\begin{tabular}{rcccccc}
  \hline
  star & $\vsini$ & $R/R_\odot$ & $2\pi R \sin i\ind{spot}/ \vsini$ & $T$ & $i( \vsini, R, T)$  & $i\ind{spot}$\\
       & (km s$^{-1}$)&          & \multicolumn{2}{c}{\dotfill\ (d) \dotfill} & \multicolumn{2}{c}{\dotfill\ ($^\circ$) \dotfill} \\
\hline
HD 49933  & 9.5$\pm$0.3 (a) & 1.39$\pm$0.02 (d)& 5.7 & 3.45$^{+0.05}_{-0.05}$ & 27$\pm$2       & 50$\pm$25 \\
HD 175726 & 13.5$\pm$0.5 (b)& 1.0$\pm$0.1  (c) & 3.2 & 3.95$^{+0.1}_{-0.1}$   & 90$^{+0}_{-30}$& 55$\pm$25 \\
HD 181420 & 18$\pm$1  (c)   & 1.6$\pm$0.1  (c) & 3.9 & 2.25$^{+0.03}_{-0.01}$ & 30$\pm$4       & 60$\pm$25 \\
HD 181906 & 10$\pm$1 (c)    & 1.4$\pm$0.1  (c) & 5.0 & 2.71$^{+0.03}_{-0.01}$ & 23$\pm$4       & 45$\pm$25 \\
  \hline
\end{tabular}
\par
References: (a) Mosser et al. (2005); (b) Kovtyukh et al. (2004); (c) Bruntt et al. (2009); (d) Goupil et al. (2009)
\end{table*}

\subsection{Inclination}

Precisely determining the star inclination has proven to be difficult, and we cannot exclude that the results are model dependent. We did not use any indication derived from the stellar $\vsini$.
Error bars for all targets are at least of $\pm 25^\circ$. This means that the fitting process is much less efficient than for the rotation period and spot lifetime.
Surprisingly, all four values we find (Table~\ref{results}) correspond to inclinations in the range [45 - 60$^\circ$]. According to the preliminary work on synthetic data, we suspect a bias in favor of such mid-values, but the inclination values derived from asteroseismology then present the same behavior, or the same bias. Such a bias could be due to the fact that rotational multiplets in the case of mid-inclination are more complex compared to the degenerate cases $i=0^\circ$ (only $\ell=0$ modes are observable) or 90$^\circ$ (only even values of $|\ell+m|$ present non-zero visibility). However, the sample of stars presently is too limited to draw conclusions.

Obtaining reliable measurements on differential rotation should be supported by other measurements, by avoiding uncertainties due to an inaccurate determination of the stellar inclination.
The best unbiased estimate of stellar inclination should come from the combined determinations of the rotation period  $\Tmoy$ from spot modeling, of the stellar radius $R$ determined from the seismic modeling, and of $v\sin i$ derived from ground-based high-resolution spectroscopy. In Table~\ref{incli}, we present  the corrected inclinations we obtain according to this method. Up to now, only HD49933 benefits from precise radius measurement derived from seismic modeling. The discrepancies in the inclination are high and deserve further work.
We should take these new values of the inclination to reestimate the differential rotation rate. However, as mentioned above, further work is needed to obtain reliable estimates and uncertainties.

\section{Conclusion\label{conclusion}}

The new data provided by CoRoT renew spot modeling with white light photometric data. With long duration, high duty-cycle, very high signal-to-noise ratio data, we can now address the case of short-lived spots in solar-like stars with solar-like activity. The only differences between the active Sun and this sample of stars are for the types (F2$\to$G0 compared to the solar type G2) and the rotation periods (2.7$\to$3.9\,d compared to a mean value of 27.3\,d).

The work presented here is a first step in addressing the study of such stars. The method we use is simpler than many previous studies,  for example \cite{2006PASP..118.1351C} analyzing MOST observations, but here, we deal with many short-lived modes, which yields a much larger number of free parameters.

The maximum entropy method proposed by \cite{2009A&A...493..193L} is capable of fitting the light curve with high accuracy because it uses a continuous distribution of spots, but it cannot follow the evolution of the active regions on timescales shorter than a single rotation period. Moreover, stellar rotation period and inclination of the rotation axis must be fixed in advance to ensure that the regularization process converges toward a unique and stable spot configuration. In view of our limited  knowledge of these stellar parameters, we decided to use the approach described in Sect. 3 to obtain information on stellar rotation rate and inclination,  as well as on the spot lifetimes, because they are shorter than or comparable to the rotation period in most of the cases. The application of the maximum entropy method is foreseen for future studies to investigate the evolution of the complexes of activities and derive a lower limit for the surface differential rotation in our targets, as shown by, e.g., Lanza et al. (2009a, 2009b).

The main result for the four \corot\ targets we studied is our reliable and precise measurement of the spot lifetimes and mean rotation.
Of course, this requires that our starspot modeling gives an accurate representation of the spots on these stars, which seems to be confirmed by the agreement with the maximum entropy results and with the analysis of the low frequency spectrum (\cite{baudin2009}).
We also find evidence of differential rotation, and show that independent measurements of the star inclination will be of great help in inferring the rate of differential rotation. Spot modeling may then achieve complementary results to asteroseismology (\cite{2004SoPh..220..169G}, \cite{2006MNRAS.369.1281B}). A dedicated analysis of differential rotation will be presented in a forthcoming work. We will also explore other ways to perform the fit of light curves with short-lived spots, such as Markov Chain Monte Carlo methods.

\begin{acknowledgements}
We thank the referees, Gordon Walker and Bryce Croll, for their advised comments that help improving the paper. \end{acknowledgements}

\end{document}